\documentclass[lettersize,journal]{IEEEtran}
\usepackage{amsmath,amsfonts}
\usepackage[ruled, linesnumbered]{algorithm2e}
\usepackage{array}
\usepackage[caption=false,font=normalsize,labelfont=sf,textfont=sf]{subfig}
\usepackage{textcomp}
\usepackage{stfloats}
\usepackage{url}
\usepackage{verbatim}
\usepackage{graphicx}
\usepackage{cite}
\usepackage{booktabs}

\usepackage{multirow}
\usepackage{makecell}
\usepackage{amsthm}
\theoremstyle{plain}

\theoremstyle{definition}

\theoremstyle{plain}

\theoremstyle{plain}

\newcommand{\sssec}[1]{\vspace*{0.05in}\noindent\textbf{#1} }

\begin{document}

\title{Throughput and Link Utilization Improvement in Satellite Networks: A  Learning-Enabled Approach}

\author{Hao Wu
\thanks{Hao Wu is with the Department of Computer Science, National University of Singapore, Singapore. (\emph{email: hao$\_$wu@nus.edu.sg})
}
}
\maketitle

\begin{abstract}
Satellite networks provide communication services to global users with an uneven geographical distribution. In densely populated regions, Inter-satellite links (ISLs) often experience congestion, blocking traffic from other links and leading to low link utilization and throughput. In such cases, delay-tolerant traffic can be withheld by moving satellites and carried to navigate congested areas, thereby mitigating link congestion in densely populated regions. Through rational store-and-forward decision-making, link utilization and throughput can be improved. Building on this foundation, this letter centers its focus on learning-based decision-making for satellite traffic. First, a link load prediction method based on topology isomorphism is proposed. Then, a Markov decision process (MDP) is formulated to model store-and-forward decision-making. To generate store-and-forward policies, we propose reinforcement learning algorithms based on value iteration and Q-Learning. Simulation results demonstrate that the proposed method improves throughput and link utilization while consuming less than 20$\%$ of the time required by constraint-based routing.
\end{abstract}
\section{Introduction}

With the Starlink system \cite{StarLink} leading the trend of satellite Internet, satellite networks are experiencing a rapid development. Many enterprises have invested in building Low Earth Orbit (LEO) and Medium Earth Orbit (MEO) satellite constellations (e.g., Iridium\cite{Iridium}, OneWeb\cite{henri2020oneweb}, and O3b\cite{O3b}), aiming to provide services to users worldwide.

\begin{figure}
\centering
\includegraphics[width=\columnwidth]{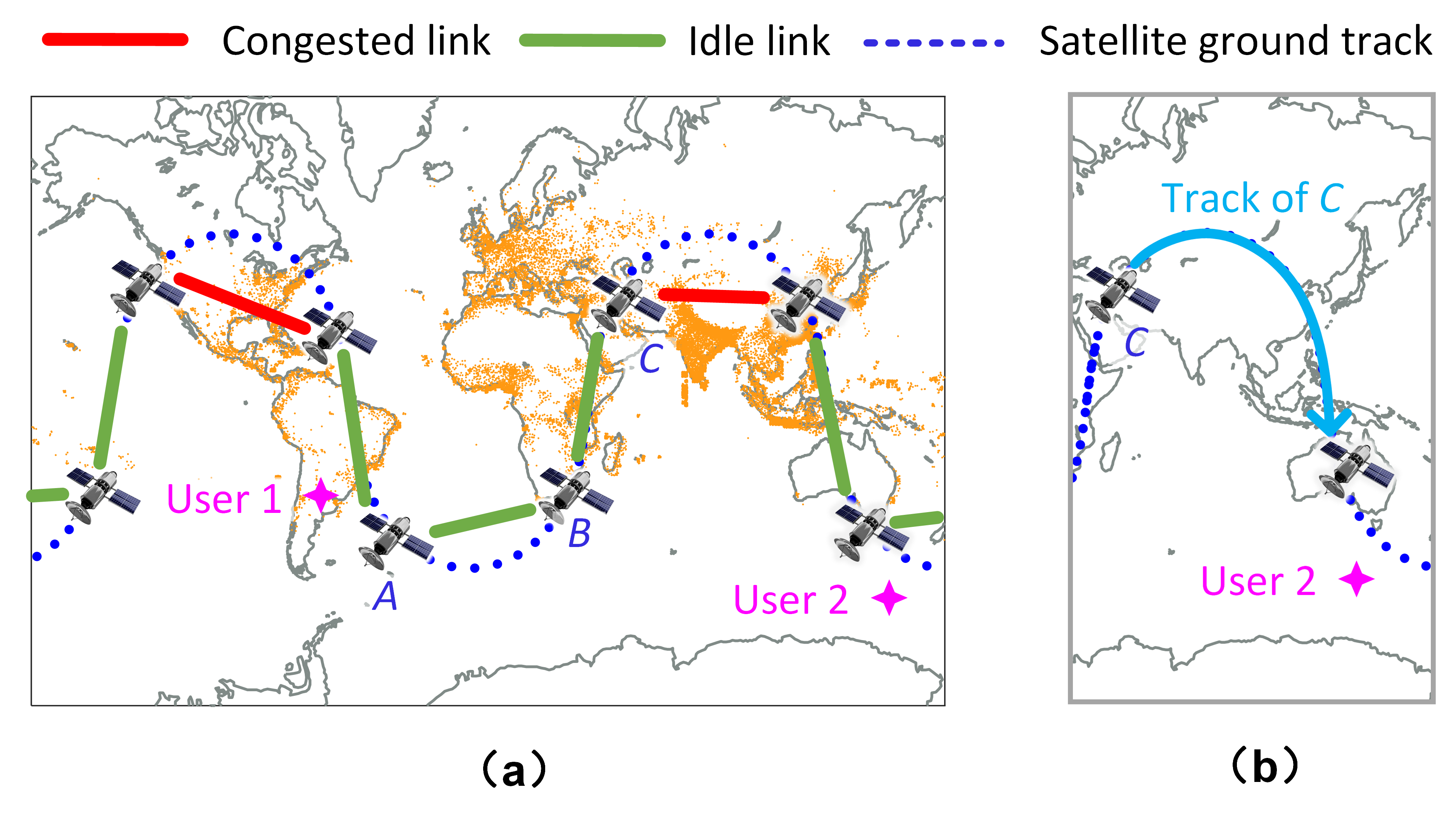}
\caption{A satellite constellation example.}
\label{fig1}
\end{figure}

Global users have an uneven geographical distribution, resulting in uneven traffic inputs to satellite networks. Consequently, ISLs in densely populated regions are typically busy or congested. End-to-end paths containing these congested ISLs cannot deliver data, although they may contain idle ISLs in sparsely populated regions. Fig. \ref{fig1}(a) shows an example of a MEO satellite constellation designed by Tsinghua University, named TSN-1A\cite{TSN}. The presence of congested links prevents the transmission of traffic between the two users, even when other links are idle, resulting in low link utilization.


To mitigate the issue of low link utilization, Internet service providers often construct networks with an asymmetrical resource layout. For instance, devices and fibers in densely populated regions, such as large cities, are increased in long-term infrastructure evolution. However, this strategy is not effective in satellite networks because the periodic orbiting of satellites equalizes the resources in different regions.

Store-and-forward methods\cite{fraire2021routing,8477064,10154792}
can be utilized to improve network performance. For instance, delay-tolerant flows between the users in Fig. \ref{fig1}(a) can be forwarded from satellite $A$ to $C$. As illustrated in Fig. \ref{fig1}(b), satellite $C$ then stores and carries the data to navigate the densely populated region until reaching User 2. This approach effectively utilizes the ISLs between $A$ and $C$. However, ensuring rational store-and-forward decisions in general cases faces two challenges.

The first challenge is to predict link loads. Store-and-forward routing approaches require the connectivity and load of each link in the future to construct a contact graph or time-expended graph (TEG). For a network with a static topology, we can predict traffic matrices using statistical methods (e.g., ARIMA \cite{601746}, wavelet transform \cite{958331}) or machine learning (e.g., recurrent neural network\cite{gao2020incorporating}, long-short term memory network\cite{8717920}) and calculate link loads. However, each user's access beam and satellite are time-varying due to the mobility of satellites. As a result, the switches in satellites must identify the end users of each packet to monitor the demands of specific flows or flow groups, which is necessary for traffic matrix prediction. Unfortunately, the implementation of sufficient flow table entries in satellite switches to support global-level traffic tracking still faces unaddressed technical issues (§II-A). 


The second challenge lies in store-and-forward decision-making for numerous flows. Traditional routing algorithms can be executed on TEGs to find store-and-forward paths. However, these algorithms typically exhibit a  computational complexity polynomially related to the scales of nodes and time slots involved. In scenarios involving long-time (e.g., 1 hour) TEGs of mega-constellations, the overall algorithm execution time might exceed the transmission deadline, making the store-and-forward decision expired.  

To address the aforementioned challenges, this paper proposes a reinforcement learning-based approach for store-and-forward decision-making. We first propose a link load prediction method based on topology isomorphism. Then, we establish a MDP model, incorporating the predicted link loads into the state transition probability and reward function. To obtain an optimal store-and-forward policy in each satellite, we design two algorithms based on value iteration and Q-Learning and evaluate their performance in a simulated system of the TSN-1A constellation. 
\section{Link Load Prediction Design}
\subsection{Infeasibility of Existing Approaches in Satellite Networks}
In satellite networks, delay-sensitive flows require an instant forward by satellites, while delay-tolerant flows can be temporarily withheld by satellites. Before scheduling the store-and-forward processes of delay-tolerant flows, the satellite network needs to predict the load of delay-sensitive flows in each link.

We represent the link load from node $i$ to $o$ at time $t$ as $\mathcal{L}_{(i,o)t}$, whose value can be expressed as a function $\mathcal{H}_t(\mathbb{T}_t)$ of traffic matrix $\mathbb{T}_t$ at $t$. The expression of $\mathcal{H}_t(\cdot)$, which depends on the traffic control mechanisms deployed, is time-varying due to network topology changes. The traffic matrix $\mathbb{T}_t$ represents the traffic demand of delay sensitive flows between each pair of satellites.  
The predicted value $\hat{\mathbb{T}}_t$ of $\mathbb{T}_t$ can be calculated based on a group of historical records $\mathbb{T}_{t-1},\mathbb{T}_{t-2},...\mathbb{T}_{0}$, using traffic matrix prediction approaches\cite{601746,958331,gao2020incorporating,8717920}. Then, the predicted link load can be obtained by solving $\hat{\mathcal{L}}_{(i,o)t} = \mathcal{H}_t(\hat{\mathbb{T}}_t)$. 

However, acquiring the historical traffic matrices in satellite networks is a considerable challenge. The mobility of satellites results in users accessing the network through different beams and satellites at different times. To accurately monitor the traffic demand between a specific user pair, satellite switches must distinguish the traffic of that user pair from other network traffic. In this context, each satellite is required to classify traffic for every user pair. However, the implementation of a satellite switch with sufficient flow table entries to support global-level traffic monitoring
has not been reported, and its feasibility is still a question. The main technical issue is the unmatch between the high energy and resource consumption of ternary content-addressable memory (TCAM)\cite{irfan2022reconfigurable} and the rigorous resource and energy constraints in satellites.

\subsection{Link Load Prediction Based on Topology Isomorphism}
\begin{figure}[thb]
\centering
\includegraphics[width=0.75\columnwidth]{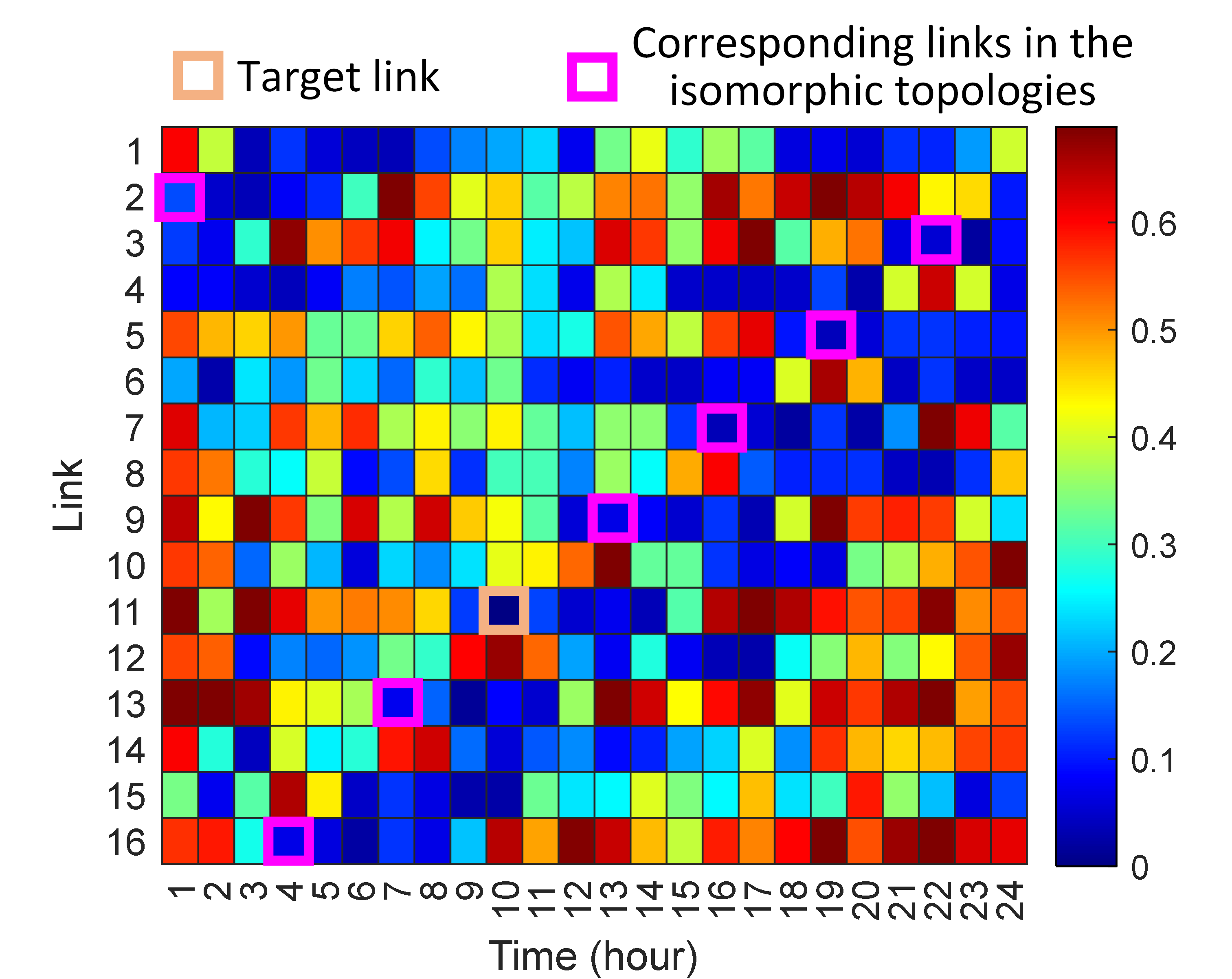}
\caption{A heatmap of Jensen-Shannon divergence.}
\label{fig2}
\end{figure}

In this letter, we propose a link load prediction method tailored for satellite networks. This method predicts the load of a given link by leveraging the statistical distribution of loads in correlated links. We first explain the correlation between link loads through a detailed analysis of experimental data obtained from the TSN-1A system. Fig. \ref{fig2} illustrates a heatmap depicting the Jensen-Shannon divergence (JSD) between the load distributions of a target link (i.e., link 11 during the $10^\text{th}$ hour) and 16 links over a 24-hour period. A lower JSD value indicates a stronger correlation between the distributions. The outcomes reveal that the target link has a higher degree of correlation with links located around the corresponding links within \emph{isomorphic topologies}. The definition of the isomorphic topology is as follows.

The topology of a satellite network at time $t$ is denoted by a directed graph $\mathcal{G}_t=(\mathcal{N}_t,\mathcal{E}_t)$, where $\mathcal{N}_t$ and $\mathcal{E}_t$ represent the node set and ISL set, respectively. The network is considered to possess isomorphic topologies between times $t$ and $t'$ if there exists a bijection $g: \mathcal{N}_t\rightarrow \mathcal{N}_{t'}$ of nodes that satisfies the following conditions. 

\begin{itemize}
	\item For any ISL existing from satellite $i$ to satellite $o$ at time $t$, there exists a corresponding ISL with an identical link rate from satellite $g(i)$ to $g(o)$ at time $t'$, and vice versa.
	\item At time $t'$, satellite $g(i)$ should cover the same geographical region as satellite $i$ does at time $t$.
\end{itemize}
Upon satisfaction of these conditions, we can establish two identical traffic control models (e.g., throughput/utility maximization) at times $t$ and $t'$. Consequently, the corresponding links within the isomorphic topologies exhibit the same distribution when the traffic intensity remains stationary. This characteristics accounts for the observed low JSD between the load distributions of the target link and its corresponding links. 

To predict the load from satellite $i$ to $o$ at time $t$, we identify a large number of times $t_1, t_2,\ldots,t_m$ before $t$ at which the corresponding bijections $g_1, g_2,\ldots,g_m$ exist. Subsequently, the distributions of link loads $\mathcal{L}_{(g_1(i),g_1(o))t_1}$, $\mathcal{L}_{(g_2(i),g_2(o))t_2}$,...$\mathcal{L}_{(g_m(i),g_m(o))t_m}$ can be employed as the predicted distribution for $\mathcal{L}_{(i,o)t}$. The approach to identifying the isomorphic topologies may vary in different satellite constellations. In some cases, isomorphism approximation can be utilized, albeit with potential performance losses due to prediction errors. For instance, the Iridium constellation, comprising six orbits inclined at $86.4^\circ$ with 11 satellites in each orbit, may be treated as having isomorphic topologies every 10 minutes by approximating the orbit inclination as $90^\circ$ and disregarding Earth's rotation in a short time. 
\section{store-and-forward Decision-Making}
\begin{figure}[thb]
\centering
\includegraphics[width=0.9\columnwidth]{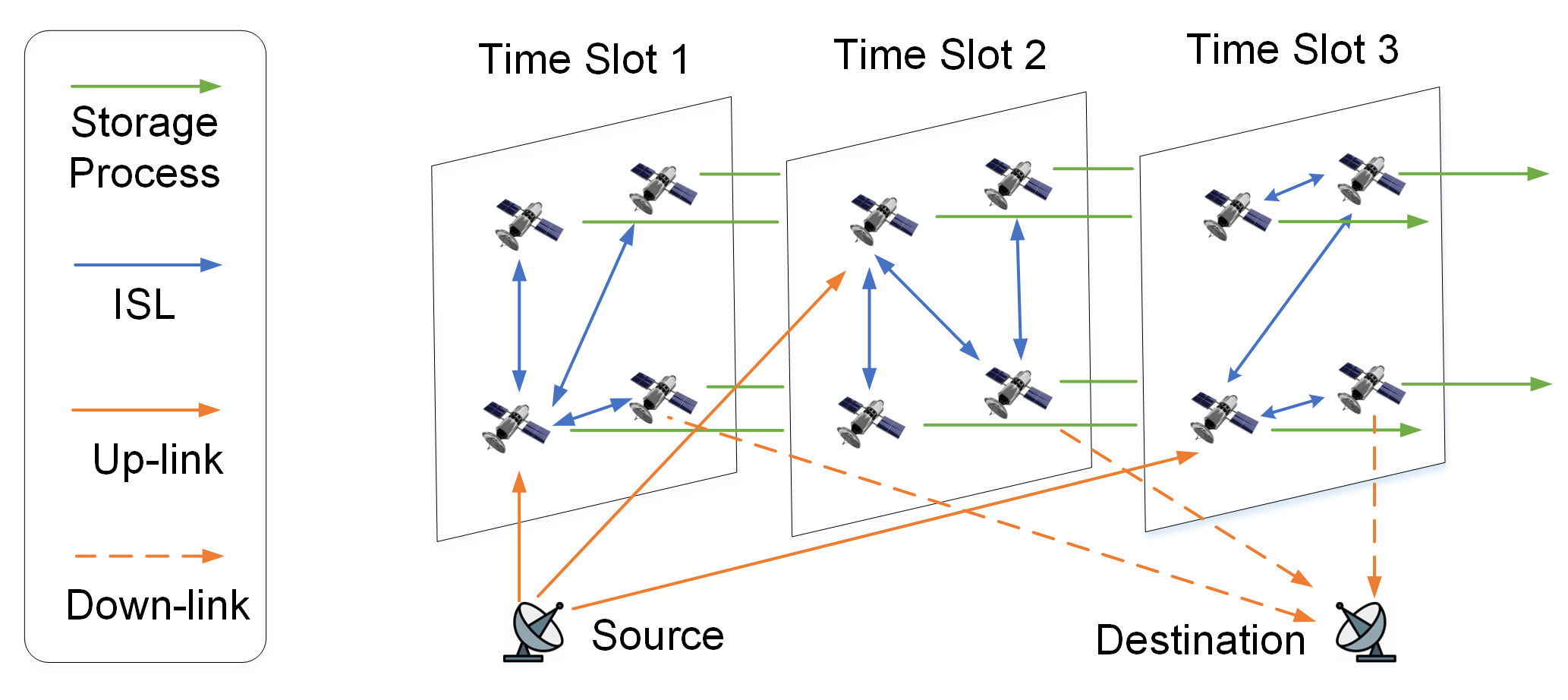}
\caption{An example of the TEG.}
\label{fig3}
\end{figure}

To generate store-and-forward policies, we employ the concept of the TEG. A TEG incorporates network snapshots captured at different time slots and introduces storage edges along with candidate up/down-links, as illustrated in Fig. \ref{fig3}. Specifically, a storage edge connects instances of a satellite within two consecutive snapshots. The capacity of a storage edge is determined as the minimum value between \emph{storage IO speed * slot length} and \emph{storage size}. 

In this scenario, constraint-based routing (CR) can be employed to select a path with sufficient bandwidth from the $k-$shortest paths, and the selected path represents a store-and-forward policy. The computational complexity of this process\footnote{$O(kn(m+n\log(n)))$ using heap structure, where $m$ is the number of edges. More information can be found in \cite{YenS}} is $O(kn^3)$. On a TEG with $n$ nodes, computing store-and-forward policies between all $O(n^2)$ source-destination pairs incurs a total complexity of $O(kn^5)$, resulting in time-consuming policy updates in mega-constellations. 

In this letter, we formulate store-and-forward decision-making as a MDP and leverage model-based value iteration and model-free Q-Learning to update policies. Compared to CR, the proposed methods have lower complexities. Specifically, value iteration requires multiple rounds of state value updates until reaching convergent store-and-forward policies, with a complexity of $O(m'n)$, where $m'$ denotes the number of iteration rounds. Q-Learning update the store-and-forward policy in each node from the previous decision, which involves at most $n$ Q-value updates with a complexity of $O(n)$.

\subsection {MDP Model}
A standard MDP model is defined by a four-tuple $(S,A,P,R)$, comprising a state space $S$, an action space $A(s)$, state transition possibilities $P(s'|s,a)$, and a reward function $R(s'|s,a)$. In the context of store-and-forward decision-making their interpretations are as follows:

\sssec{State space $S$:} The state $s$ of a packet represents a node in the TEG (i.e., the source, destination, or a satellite instance) that holds the packet. In a TEG denoted by a directed graph $\mathcal{G}=\{\mathcal{N},\mathcal{E}\}$, the value of $s$ is the node index, ranging from 1 to $|\mathcal{N}|$. 
Additionally, a packet drop state with index $|\mathcal{N}|+1$ is introduced. Thus, the state space is defined as $S=\{1,2,...,|\mathcal{N}|+1\}$.

\sssec{Action space $A$:} The action $a$ for a packet involves forwarding it to a specific node or storing it in the current node, with the value of $a$ representing the index of the next node. Consequently, the action space in state $s$ is expressed as $A_s = \{a | (s,a)\in\mathcal{E}\}$.
Specifically, $A_{|\mathcal{N}|+1} = \{|\mathcal{N}|+1\}$.
\begin{algorithm}[t]
\caption{Model Element Calculation}\label{Rewards}
\KwData{link load prediction result $p(\mathcal{L}_{(i,o)})$, $C_{(i,o)}$, $\mathcal{L}'_{(i,o)}$,  $c_0,c'_0,c''_0,c_1,c'_1,\epsilon$}
\KwResult{$r(s,a),P(s'|s,a)$}
\For {$s\in S,a\in A_s$}{
	\uIf (\tcp*[h] {packet drop}){$s = |\mathcal{N}|+1$}{
	$r(s,a) \leftarrow c''_0 $\;
	$P(|\mathcal{N}|+1|s,a) \leftarrow 1$;}
	\uElseIf{$(s,a)$ is a storage edge}
		{\uIf (\tcp*[h] {succeed}){$\mathcal{L}'_{(s,a)}> C_{(s,a)}$}
		{$r(s,a)\leftarrow c'_0+\frac{c'_1}{C_{(s,a)}-\mathcal{L}'_{(s,a)}}$\;
		$P(a|s,a) \leftarrow 1$;}
		\lElse(\tcp*[h] {fail}){remove $a$ from $A_s$}}
	\Else(\tcp*[h] {forwarding})
		{$P(a|s,a)=\int_0^{\mathcal{C}_{(s,a)}-\mathcal{L}'_{(s,a)}}p(\mathcal{L}_{(s,a)})d\mathcal{L}_{(s,a)}$\;
        \lIf{$P(a|s,a)=0$}{remove $a$ from $A_s$}
		Let $s^+$ satisfies that $(s,s^+)$ represents storage\;
		\lIf{$\mathcal{L}'_{(s, s^+)}>\mathcal{C}_{(s,s^+)}$}{$P(s^+|s,a)=0$}
		\lElse{$P(s^+|s,a)=1-P(a|s,a)$}
		$P(|\mathcal{N}|+1\}|s,a)\leftarrow 1-P(a|s,a)-P(s^+|s,a)$\;
		$r(s,a) \leftarrow \sum_{s'= a,s^+,|\mathcal{N}|+1}P(s'|s,a)R(s'|s,a)$
		}
}
\Return $r(s,a),P(s'|s,a)$ 
\end{algorithm}

\sssec{State transition probabilities $P(s'|s,a)$:} 
The edge $(s,a)\in\mathcal{E}$ on the TEG $\mathcal{G}=\{\mathcal{N},\mathcal{E}\}$ represents either the forwarding process or storage process from node $s$ to $a$. The capacity corresponding to $(s,a)$ is denoted by $\mathcal{C}_{(s,a)}$, and $\mathcal{L}'_{(s,a)}$ represents the bandwidth or storage space already allocated to delay-tolerant flows.
In the case of a storage process, where $\mathcal{L}_{(s,a)}$ is 0 due to the absence of stored delay-sensitive flows, successful storage (i.e., $P(a|s,a)=1$) occurs if $\mathcal{C}_{(s,a)}>\mathcal{L}'_{(s,a)}$. Otherwise, storage fails, leading to packet loss (i.e., $P(|\mathcal{N}|+1|s,a)=1$).

For a forwarding process, the possibility of success is determined by $P(\mathcal{C}_{(s,a)}-\mathcal{L}'_{(s,a)}>\mathcal{L}_{(s,a)})$, which is calculated based on the predicted distribution of link load $\mathcal{L}_{(s,a)}$. If satellite $s$ still has available storage space, then the packet is temporarily stored with a probability of $1-P(\mathcal{C}_{(s,a)}-\mathcal{L}'_{(s,a)}>\mathcal{L}_{(s,a)})$; otherwise, the packet has to be dropped.    

\sssec{Reward function $R(s'|s,a)$.} This function is designed to facilitate a rational policy that prevents the overuse of high-load links and on-board storage spaces. Given a reward function, the expected reward of taking action $a$ at state $s$ is:
\begin{equation}
r(s,a) = \sum\nolimits_{s'=a,s^+,|\mathcal{N}|+1}P(s'|s,a)R(s'|s,a),
\end{equation}
where the value of $R(s'|s,a)$ is assigned as:
\begin{equation}
    \begin{aligned}
    & R(a|s,a) = c_0 + c_1(\mathcal{C}_{(s,a)}-\mathcal{L}'_{(s,a)}-\mathcal{L}_{(s,a)})^{\text{-1}},\\
    & R(s^+|s,a) = c'_0 + c'_1(\mathcal{C}_{(s,a)}-\mathcal{L}'_{(s,a)})^{\text{-1}},\\
    & R(|\mathcal{N}|+1|s,a) = c''_0,
    \end{aligned}
\label{reward_value}
\end{equation}
and $c_0, c'_0, c''_0, c_1, c'_1$ are tunnable cost coefficients. 
Before policy update, the network updates the reward and state transition possibilities as in Algorithm 1. 

\subsection{Policy Generation based on Value Iteration}
The network employs the standard value iteration algorithm, iterating as (\ref{VE}) for every state $s\in S$ except the destination $n_{des}$ until all the values converge. 
\begin{equation}
\centering
V(s)\leftarrow \max\limits_{a\in A_s)}r(s,a)+\gamma\sum_{s'}P(s'|s,a)V(s).
\label{VE}
\end{equation}
Then, the per-node store-and-forward policies $\pi(s),s\in S$ for a specific source-destination pair is generated as follows.
\begin{algorithm}
\caption{Policy Generation from $n_{src}$ to $n_{des}$}
\KwData{$(S,A,P,R)$, $r(s,a)$, $V(s)$}
\KwResult{$\pi(s)$}
$n\leftarrow n_{src}$\;
$l\leftarrow \text{a list with one element $n$}$\;
\While{$n_{des}\notin l$ and  $|\mathcal{N}|+1\notin l$}
{$\pi(n)\leftarrow \arg\max\limits_{a\in A_n\cap\overline{l}}r(n,a)+\gamma\sum_{s'}P(s'|n,a)V(s')$\;
$n\leftarrow \pi(n)$;
}
\Return $\pi(s)$ 
\end{algorithm}

\begin{figure}[th]
\centering
\includegraphics[width=0.6\columnwidth]{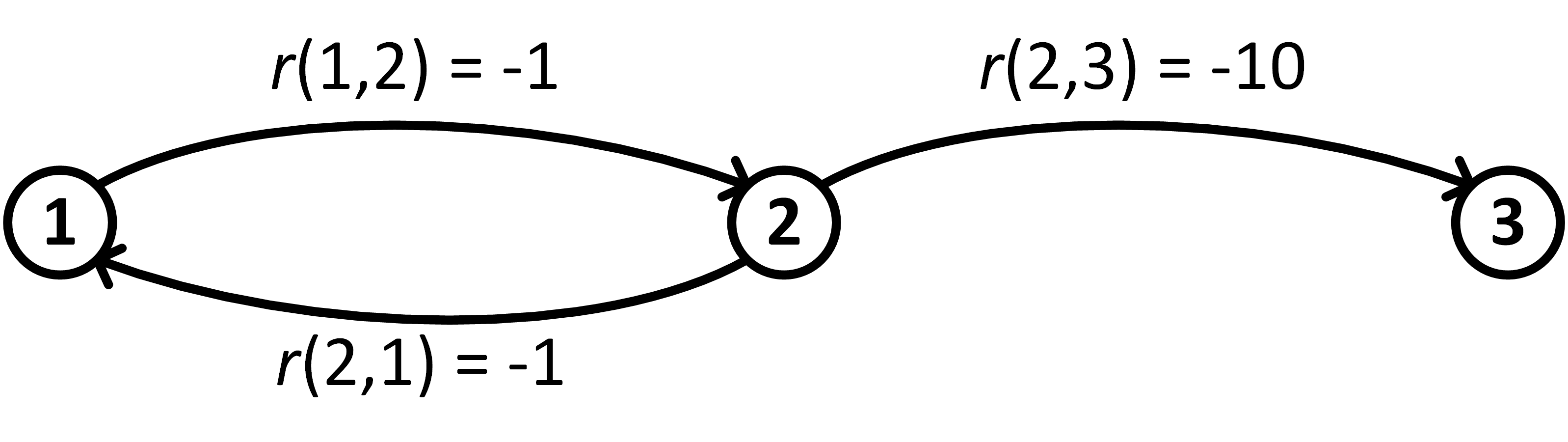}
\caption{An example of routing loop.}
\label{fig4}
\end{figure}

The introduction of a discount factor $\gamma < 1$ can potentially lead to routing loop problems if each node directly chooses the max-reward policy as 
\begin{equation}
\pi(n)\leftarrow \arg\max\limits_{a\in A_n}r(n,a)+\gamma\sum_{s'}P(s'|n,a)V(s').
\end{equation}
In Fig. \ref{fig4}, an example illustrates the network generating policies from node 1 to 3 with $\gamma=0.5$. The total reward along the looped path between 1 and 2 is $r(2,1)+\gamma r(1,2)+\gamma^2 r(2,1)...=-1(1+\gamma+\gamma^2+\ldots+\gamma^\infty)=-2$, which surpasses $r(2,3)$. Consequently, node 2 will forward the traffic back to node 1. This issue is not easily resolved by adjusting $\gamma$. Therefore, we modify the feasible region of action $a$ to $A_n\cap\overline{l}$ to avoid revisiting any node already in the path $l$. 

\subsection{Policy Generation based on Q-Learning}
Q-Learning is an model-free reinforcement learning method that learns policies from previous decisions. To balance exploitation and exploration of new paths, an $\epsilon-$greedy strategy is employed. By using Algorithm 3, the network updates Q-values for each source-destination pair, which skips the procedure of computing $V(s)$. As the update of $Q(s,a)$ occurs, the policies gradually converge over time.

\subsection{Fairness among Source-Destination Pairs}
As flows occupy more bandwidth, the values of $\mathcal{L}'_{(i,o)}$ increase. Consequently, the network requires to update $P(s'|s,a)$, $r(s,a)$, and the accompanying policies using the algorithms described above. In this context, we control the policy update frequency through a \emph{maximum reserved data volume} denoted as $\mathcal{B}$, which is tuned before use (§IV-B). The network employs a round-robin mechanism to schedule flows between different source-destination pairs. During each round, the network can allocate a maximum of $\mathcal{B}$ data for flows between each source-destination pair. At the end of each round, the network executes Algorithm 1 along with one of the policy generation approaches to update polices.
\begin{algorithm}
\caption{Policy Generation from $n_{src}$ to $n_{des}$}
\KwData{$(S,A,P,R)$, $r(s,a)$}
\KwResult{$\pi(s)$}
$n\leftarrow n_{src}, l\leftarrow \text{a list with element $n$}$\;
$Q(s,a)\leftarrow 0,\forall s,a$\;
\While{$n_{des}\notin l$ and  $|\mathcal{N}|+1\notin l$}
{Generate a uniform random value $c_{rdm}$ in $[0,1]$\;
\lIf{$c_{rdm}>\epsilon$}{$n'\leftarrow \arg\max_{a\in A(n)\cap \overline{l}}Q(n,a)$} 
\lElse{$n'\leftarrow$ a random element in $A(n)\cap \overline{l}$}
Add $n'$ at the end of $l$\;
$Q(n,n')\leftarrow Q(n,n')+\alpha\left(r(n,n')+\gamma\left(\max_{a\in A(n')\cap \overline{l}}Q(n',a)-Q(n,n')\right)\right)$;\
$\pi(n)\leftarrow\arg\max_{a\in A(n)\cap \overline{l}}Q(n,a)$;
}
\Return $\pi(s)$ 
\end{algorithm}
\section{Simulation}
\subsection{Simulation Setup}
The experiments are conducted on a simulated constellation of TSN-1A\cite{TSN}, which comprises eight circular orbits with one MEO satellite in each orbit. Each satellite is connected to two adjacent satellites through two 10 Gbps links and communicates with terminals through 8 Ka-band beams\footnote{To expedite the simulation, all links have a 20-fold reduction on capacity.}. 
The simulation also involves the deployment of 970 small terminals, 390 large terminals, and 6 ground stations, whose distribution is illustrated in Fig. \ref{fig5}(a). They generate delay-sensitive flows, encompassing voice, video, and file transmissions. The remaining bandwidth is allocated for transmitting delay-tolerant data move flows among 3 ground stations situated in Shanghai, Frankfurt, and Toronto. A key performance metric recorded in the simulation is the total throughput of delay-tolerant flows.
\subsection{Parameter Tuning}
\begin{figure*}
\centering
\subfloat[]{\includegraphics[height=1.2in]{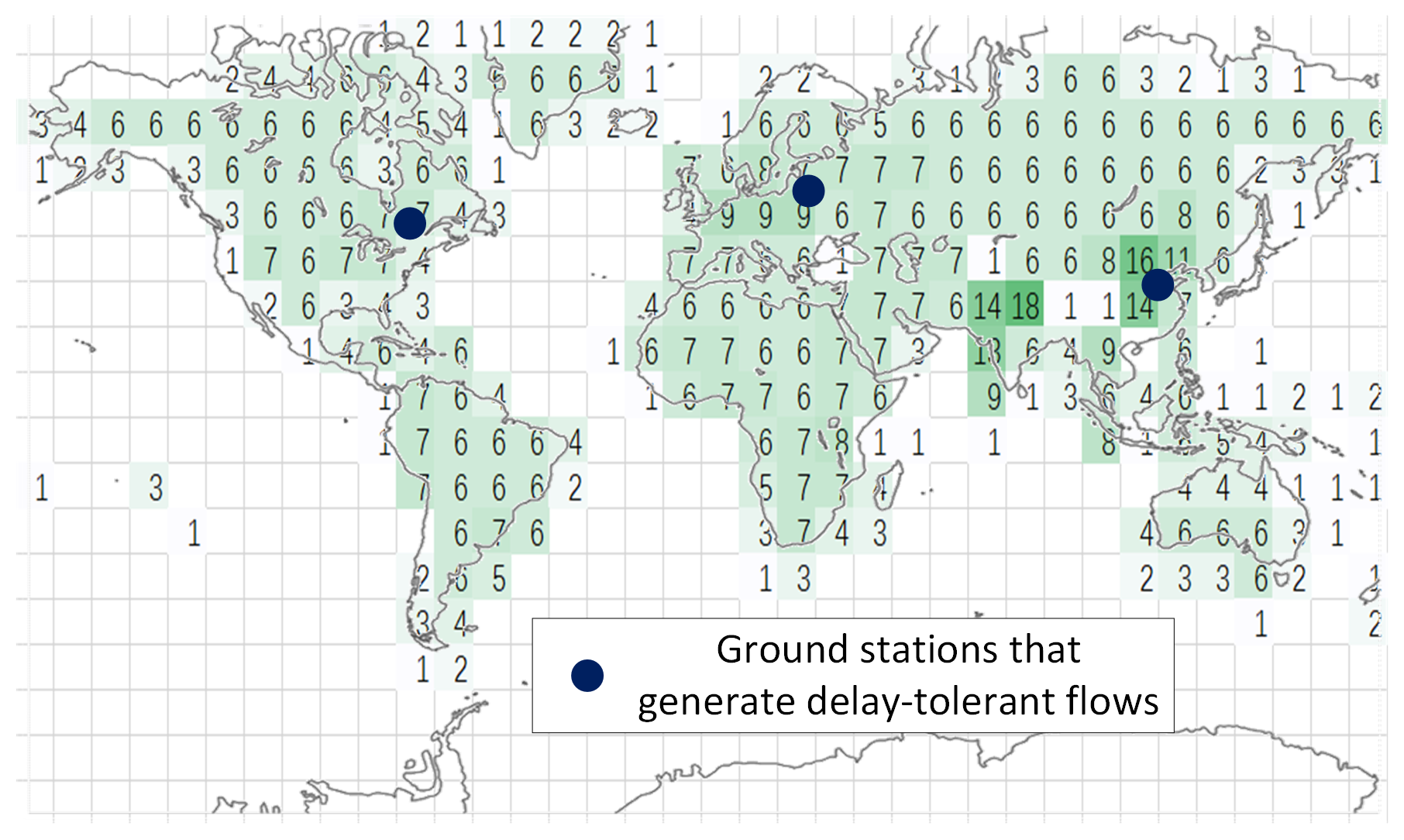}}\quad
\subfloat[]{\includegraphics[height=1.2in]{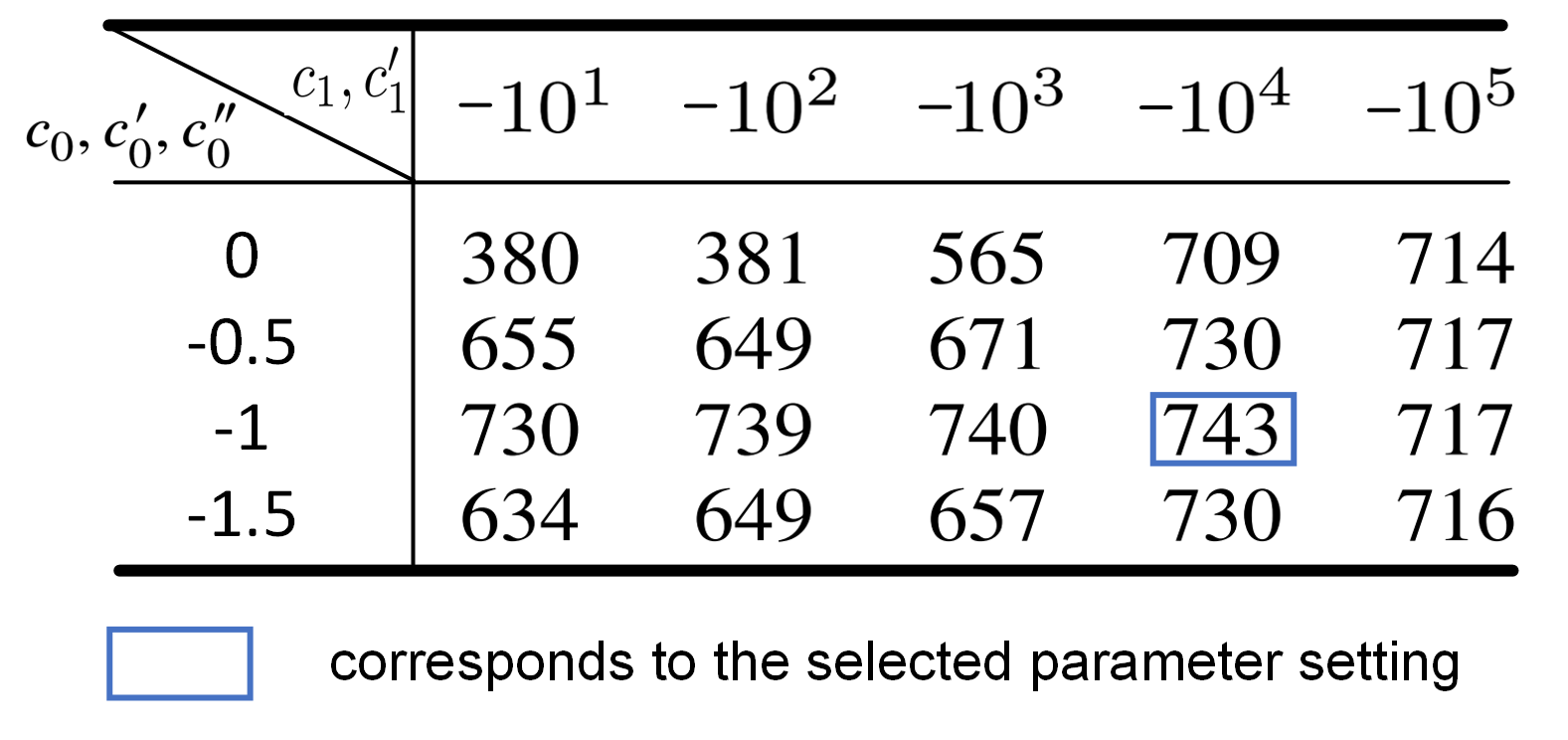}}\quad
\subfloat[]{\includegraphics[height=1.2in]{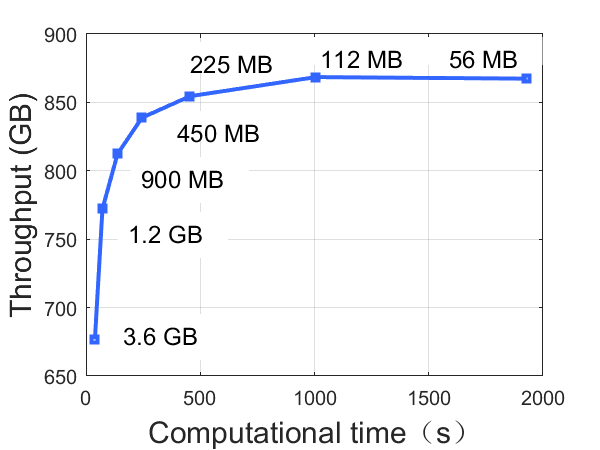}}
\caption{Simulation setup and parameter tuning.}
\label{fig5}
\end{figure*}

The cost coefficients $c_0,c'_0,c''_0, c_1,c'_1$ in the reward function (\ref{reward_value}) influence the performance of decision-making. Therefore, a tuning of their values is conducted, considering different on-board storage sizes. For instance, Fig. \ref{fig5}(b)  presents the 24-hour throughput with a storage size of 400 GB. Parameter settings yielding the maximum throughput are selected.

As mentioned in Section III-D, the maximum reserved data volume $\mathcal{B}$ also requires tuning. As illustrated in Fig. \ref{fig5}(c), the performance curve of decision-making over 24 hours using Q-Learning is presented, showing the impact of $\mathcal{B}$. It is observed that as $\mathcal{B}$ increases, both throughput and computational time grow. Then, the tuning of $\mathcal{B}$ (e.g., choosing 225 MB) ensures high throughput while maintaining a relatively short time.

\begin{figure*}
\centering
\subfloat[]{\includegraphics[height=1.3in]{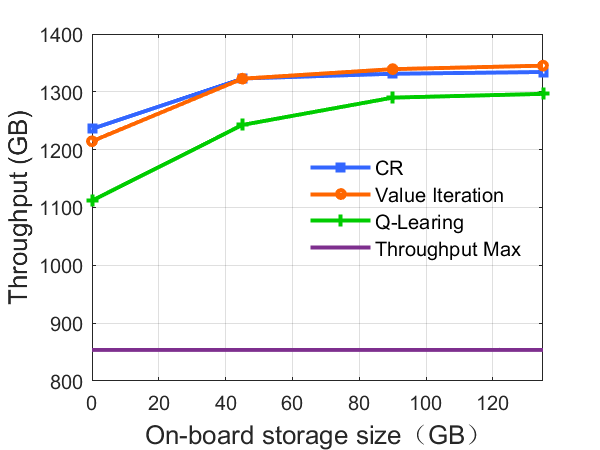}}
\subfloat[]{\includegraphics[height=1.3in]{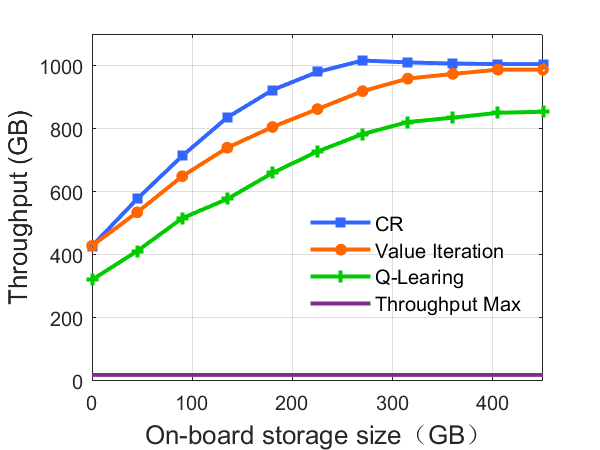}}
\subfloat[]{\includegraphics[height=1.3in]{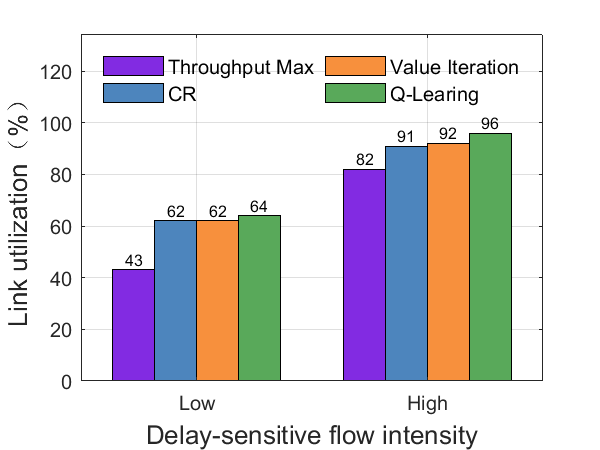}}
\subfloat[]{\includegraphics[height=1.3in]{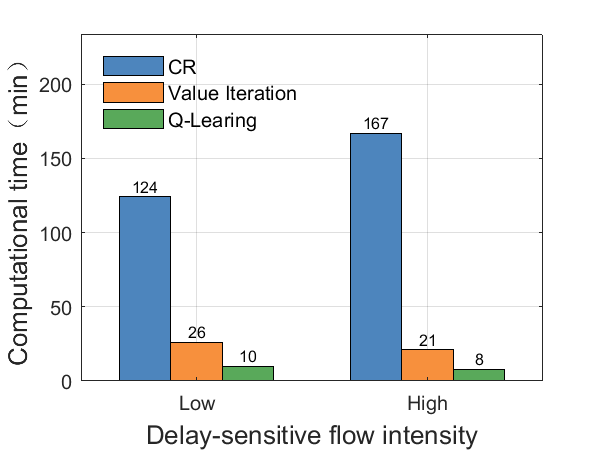}}
\caption{Performance evaluation. (a) Throughput under a low delay-sensitive flow intensity. (b) Throughput under a high delay-sensitive flow intensity. (c) Link utilization. (d) Computational time.}
\label{fig6}
\end{figure*}
\subsection{Performance Evaluation}
We setup two scenarios with low and high delay-sensitive flow intensities, where 56 $\%$ and 18 $\%$ link capacity on average remains for delay-tolerant flows, respectively. In these scenarios, we conduct a comparative analysis of the proposed algorithms against a standard throughput maximization solver without on-board storage and CR with on-board storage. The results, as depicted in Fig. \ref{fig6}(a)(b), reveal that the algorithm based on value iteration achieves a throughput close to CR, while the algorithm based on Q-Learning obtains slightly lower throughput. In addition, with the growth of on-board storage size grows, the performance of these approaches saturates. Furthermore, the proposed algorithms contribute to an improvement in link utilization, as depicted in Fig. \ref{fig6}(c). 

Fig. \ref{fig6}(d) illustrates the total computation time of policy updates over 24 hours using the same CPU (Intel Core i7-8700 CPU @3.20GHz). Both the proposed algorithms consume less than 20$\%$ of the time required by the CR algorithm. More importantly, as discussed in Section III, the computational complexity of the proposed algorithms grows much more slowly than the complexity of CR as the TEG scales. Therefore, a more substantial reduction in computation time can be anticipated in mega-constellations with thousands of satellites.

\section{conclusion}
This letter proposes a learning-enabled store-and-forward decision-making method for satellite traffic. This method first predicts the traffic loads on ISLs by utilizing topology isomorphism. Then, the store-and-forward decision-making is modelled as a MDP, and two policy generation algorithms based on value iteration and Q-Learning are proposed. Simulation results demonstrate that the proposed algorithms can improve throughput and link utilization while consuming less than 20$\%$ of the time required by the CR algorithm. 

\ifCLASSOPTIONcaptionsoff
  \newpage
\fi
\bibliographystyle{IEEEtran}
\bibliography{reference.bib}

\begin{thebibliography}{10}
\providecommand{\url}[1]{#1}
\csname url@samestyle\endcsname
\providecommand{\newblock}{\relax}
\providecommand{\bibinfo}[2]{#2}
\providecommand{\BIBentrySTDinterwordspacing}{\spaceskip=0pt\relax}
\providecommand{\BIBentryALTinterwordstretchfactor}{4}
\providecommand{\BIBentryALTinterwordspacing}{\spaceskip=\fontdimen2\font plus
\BIBentryALTinterwordstretchfactor\fontdimen3\font minus
  \fontdimen4\font\relax}
\providecommand{\BIBforeignlanguage}[2]{{%
\expandafter\ifx\csname l@#1\endcsname\relax
\typeout{** WARNING: IEEEtran.bst: No hyphenation pattern has been}%
\typeout{** loaded for the language `#1'. Using the pattern for}%
\typeout{** the default language instead.}%
\else
\language=\csname l@#1\endcsname
\fi
#2}}
\providecommand{\BIBdecl}{\relax}
\BIBdecl

\bibitem{StarLink}
\BIBentryALTinterwordspacing
SpaceX. (2024) Starlink. [Online]. Available: \url{https://www.spacex.com/}
\BIBentrySTDinterwordspacing

\bibitem{Iridium}
\BIBentryALTinterwordspacing
Iridium. (2024) Iridium. [Online]. Available: \url{https://www.iridium.com/}
\BIBentrySTDinterwordspacing

\bibitem{henri2020oneweb}
Y.~Henri, ``The oneweb satellite system,'' in \emph{Handbook of Small
  Satellites: Technology, Design, Manufacture, Applications, Economics and
  Regulation}.\hskip 1em plus 0.5em minus 0.4em\relax Springer, 2020, pp.
  1091--1100.

\bibitem{O3b}
\BIBentryALTinterwordspacing
O3b. (2024) Proven high-performance connectivity. [Online]. Available:
  \url{https://www.ses.com/our-coverage/o3b-meo}
\BIBentrySTDinterwordspacing

\bibitem{TSN}
ITU-R, ``Coordination of the tsn-1a satellite network in br ific 2894,''
  \url{http://www.itu.int/sns/ific8/ific2894.zip}, Online database, 2023.

\bibitem{fraire2021routing}
J.~A. Fraire, O.~De~Jonck{\`e}re, and S.~C. Burleigh, ``Routing in the space
  internet: A contact graph routing tutorial,'' \emph{Journal of Network and
  Computer Applications}, vol. 174, p. 102884, 2021.

\bibitem{8477064}
P.~Wang, X.~Zhang, S.~Zhang, H.~Li, and T.~Zhang, ``Time-expanded graph-based
  resource allocation over the satellite networks,'' \emph{IEEE Wireless
  Communications Letters}, vol.~8, no.~2, pp. 360--363, 2019.

\bibitem{10154792}
H.~Wu, J.~Yan, and L.~Kuang, ``Utility maximization in satellite networks using
  onboard caching,'' in \emph{2023 4th Information Communication Technologies
  Conference (ICTC)}, 2023, pp. 94--98.

\bibitem{601746}
A.~Adas, ``Traffic models in broadband networks,'' \emph{IEEE Communications
  Magazine}, vol.~35, no.~7, pp. 82--89, 1997.

\bibitem{958331}
S.~Ma and C.~Ji, ``Modeling heterogeneous network traffic in wavelet domain,''
  \emph{IEEE/ACM Transactions on Networking}, vol.~9, no.~5, pp. 634--649,
  2001.

\bibitem{gao2020incorporating}
K.~Gao, D.~Li, L.~Chen, J.~Geng, F.~Gui, Y.~Cheng, and Y.~Gu, ``Incorporating
  intra-flow dependencies and inter-flow correlations for traffic matrix
  prediction,'' in \emph{2020 IEEE/ACM 28th International Symposium on Quality
  of Service (IWQoS)}.\hskip 1em plus 0.5em minus 0.4em\relax IEEE, 2020, pp.
  1--10.

\bibitem{8717920}
A.~Lazaris and V.~K. Prasanna, ``An lstm framework for modeling network
  traffic,'' in \emph{2019 IFIP/IEEE Symposium on Integrated Network and
  Service Management (IM)}, 2019, pp. 19--24.

\bibitem{irfan2022reconfigurable}
M.~Irfan, A.~I. Sanka, Z.~Ullah, and R.~C. Cheung, ``Reconfigurable
  content-addressable memory (cam) on fpgas: A tutorial and survey,''
  \emph{Future Generation Computer Systems}, vol. 128, pp. 451--465, 2022.

\bibitem{YenS}
\BIBentryALTinterwordspacing
J.~Y. Yen, ``Finding the k shortest loopless paths in a network,''
  \emph{Management Science}, vol.~17, no.~11, pp. 712--716, 1971. [Online].
  Available: \url{https://doi.org/10.1287/mnsc.17.11.712}
\BIBentrySTDinterwordspacing

\end{thebibliography}
\end{document}